\documentclass[nofootinbib,reprint,amssymb,prd,aps,superscriptaddress]{revtex4-1}
\usepackage{graphicx}
\usepackage{xcolor,amsmath}
\usepackage{multirow}
\usepackage{comment}
\usepackage{enumitem}
\usepackage{hyperref}
\hypersetup{colorlinks, linkcolor={teal},citecolor={blue},urlcolor={blue}}

\newcommand{\dd}{{\rm d}}

\begin{document}

\title{Gravitational perturbations of nonlocal black holes}

\author{Rocco D'Agostino}
\email{rocco.dagostino@inaf.it}
\affiliation{INAF--Osservatorio Astronomico di Roma, Via Frascati 33, 00078 Monte Porzio Catone, Italy}
\affiliation{INFN, Sezione di Roma 1, Piazzale Aldo Moro 2, 00185 Roma, Italy}

\author{Vittorio~De~Falco}
\email{v.defalco@ssmeridionale.it}
\affiliation{Ministero dell'Istruzione e del Merito (M.I.M., ex M.I.U.R.), Italy}

\begin{abstract}
We derive the master equations governing axial and polar gravitational perturbations of a generic static and spherically symmetric black hole spacetime within the framework of the revised Deser--Woodard nonlocal gravity theory. We then apply our general formalism to a one-parameter family of black hole solutions recently obtained by the present authors, representing small first-order deviations from the Schwarzschild geometry. We provide well-motivated arguments that allow us to render the analysis analytically tractable.
Our results provide the first complete perturbative characterization of nonlocal black holes and lay the groundwork for future investigations.

\end{abstract}

\maketitle

\section{Introduction}

Even though recent empirical advances, driven by black hole (BH) observations \cite{EventHorizonTelescope:2019dse,EventHorizonTelescope:2022wkp} and gravitational wave (GW) detections \cite{LIGOScientific:2016aoc,LIGOScientific:2017vwq,Abac:2025saz}, have further corroborated the validity of standard general relativity (GR) in different field regimes, fundamental issues remain unresolved \cite{Berti:2015itd,Barack:2018yly}. 
The primary obstacle stems from the fundamental incompatibility of GR with a quantum-level formulation, which continues to present a persistent challenge to achieving a complete comprehension of the gravitational interaction. 
This issue sharply distinguishes gravity from the other fundamental interactions, all of which are consistently described within the framework of quantum field theory \cite{Carlip:2001wq,Ashtekar:2004eh}. This theoretical divide becomes particularly acute in extreme physical domains, such as near singularities and at Planck-scale energies, where the effects of quantum gravity are expected to dominate.

Phenomenologically, Einstein's theory proves inadequate for a comprehensive explanation of the Universe's evolution. For instance, the necessity of additional components to explain the galactic dynamics, such as dark matter, suggests that GR may represent an incomplete description of gravity \cite{Sanders:2002pf,Bertone:2004pz}. An even compelling limitation of GR, however, is the required introduction of the cosmological constant, $\Lambda$, in the standard Hilbert-Einstein action to account for the observed accelerated expansion of the cosmos  \cite{SupernovaCosmologyProject:1998vns,SupernovaSearchTeam:1998fmf,Peebles:2002gy,Frieman:2008sn,DAgostino:2019wko}. Indeed, interpreting $\Lambda$ through the lens of quantum field theory gives rise to the notorious fine-tuning problem \cite{Weinberg:1988cp,Padmanabhan:2002ji,DAgostino:2022fcx}. 

The necessity of tackling these issues has resulted in the formulation of modified gravitational models \cite{Ferraro:2006jd,Sotiriou:2008rp,Clifton:2011jh,Joyce:2014kja,Koyama:2015vza,Nojiri:2017ncd,Capozziello:2019cav,DAgostino:2019hvh,DAgostino:2022tdk,DAgostino:2025kme}. 
In particular, nonlocal theories of gravity have gained attention as a highly promising direction for resolving GR inconsistencies at the fundamental level \cite{Arkani-Hamed:2002ukf,Deser:2007jk,Nojiri:2007uq,Hehl:2008eu,Deffayet:2009ca,Maggiore:2013mea,Bombacigno:2024lud}.
The rationale for nonlocal modifications of GR originates from several theoretical advantages, including their intrinsic capacity to account for cosmic acceleration \cite{Maggiore:2014sia,Dirian:2014ara,Capozziello:2022rac,Capozziello:2023ccw},
their potential to mitigate singularity problems \cite{Biswas:2011ar,Buoninfante:2018mre}, and their ability to describe gravity within quantum standards, by circumventing the conventional drawbacks of local higher-derivative formulations \cite{Calcagni:2007ru,Modesto:2011kw,Modesto:2017sdr,Boos:2022biz}. 
Crucially, these models yield intriguing and observable phenomenological signatures that are sensitive to empirical verification through forthcoming gravitational and cosmological measurements, thereby outlining a plausible route beyond standard gravitational physics.

It is also worth noting that nonlocal structures naturally emerge in a broader theoretical context. In particular, nonlocal effective actions may arise in scalar field theories when vacuum configurations are considered and certain degrees of freedom are integrated out, leading to effective Lagrangians with nonlocal kernels. Such constructions provide additional motivation for nonlocal gravity, as they suggest that nonlocal terms may be interpreted as effective descriptions of underlying microscopic dynamics rather than \textit{ad hoc} modifications
\cite{Oikonomou:2025htz}.

Among the most promising formulations within the realm of nonlocal gravity there is the Deser-Woodard (DW) model~\cite{Deser:2007jk}, which aims to heal the fine-tuning problem through the addition of a nonlocal function of the Ricci scalar in the Einstein-Hilbert action. This modification is designed to alter gravitational behaviour over large cosmic distances, while simultaneously preserving consistency with well-tested predictions of GR at smaller scales. 
The DW model does not invoke additional degrees of freedom, making it a conceptually more direct extension of GR than many other modified gravity theories.
To ensure a theoretically consistent and physically reliable framework, the original DW model has undergone a subsequent revision~\cite{Deser:2019lmm}, with the nonlocal function constrained to reproduce the observed cosmic evolution while remaining compatible with solar system tests.

Later developments of the DW model explored how nonlocal terms can also reproduce MOND-like phenomenology in gravitationally bound systems.
In particular, inverse d'Alembertian kernels acting on curvature scalars modify the Einstein equations to yield flat rotation curves and the baryonic Tully-Fisher relation without particle dark matter~\cite{Woodard:2014wia}.
Additionally, it has been emphasized that abandoning dark matter in a generally covariant, metric‑based theory inevitably leads to nonlocality, and exhibited models that reproduce the principal cosmological successes of cold dark matter~\cite{Deffayet:2024ciu}.
More recent work has constructed nonlocal frameworks that explicitly interpolate between MOND‑like dynamics in bound systems and an effective dark‑matter description at cosmological scales, unifying both regimes within a single action~\cite{Deffayet:2025lwl}.
However, although nonlocal gravity can mimic MOND effects, a complete theory must also account for galaxy spin and stability, the cosmic microwave background spectrum, baryon acoustic oscillations, and the correlation among the latter. Nonlocal frameworks can address some of these aspects, but reproducing all cosmological and astrophysical observations, including weak and strong gravitational lensing, still points to the need for particle dark matter (see, e.g., Ref.~\cite{Oikonomou:2025bsi}).

The revised DW model has been extensively applied across a variety of scenarios \cite{Ding:2019rlp,Chen:2019wlu,Jackson:2021mgw,Jackson:2023faq,DAgostino:2025sta}.
In particular, we recently considered this framework to derive a new class of BH solutions characterized by linear perturbations to the Schwarzschild spacetime. Hereafter, we refer to it as the D'Agostino--De Falco (DD) BH \cite{DAgostino:2025wgl}.
Perturbations of BHs occupy a central role in this context, as they provide one of the most sensitive arenas in which potential deviations from GR may become observable \cite{Dreyer:2003bv,Sasaki:2003xr,Barack:2018yvs}. In particular, the propagation of linearized fluctuations around compact objects encodes detailed information about the underlying gravitational geometry, making perturbation theory a powerful probe of extended frameworks \cite{Yunes:2007ss,Cardoso:2009pk,deRham:2019gha,Langlois:2021aji,DeSimone:2025sgu}. The characteristic oscillation spectrum of BHs, as well as the associated scattering and stability properties, directly influence GW signals, detectable by current and upcoming detectors \cite{Kokkotas:1999bd,Berti:2009kk,Konoplya:2011qq}. Consequently, even subtle departures from the Schwarzschild predictions arising, e.g., from effective nonlocal interactions, may imprint distinguishable signatures in ringdown waveforms or quasi-normal mode (QNM) frequencies. As GW precision continues to improve, especially with the advent of next-generation facilities, such as LISA and the Einstein Telescope \cite{ET:2025xjr,Bonilla:2019mbm,Califano:2023aji,DAgostino:2023tgm}, perturbative analyses of modified gravity models become increasingly critical. They not only enable stringent tests of the strong-field regime, but also offer a concrete pathway to identifying or constraining new physics beyond GR through astrophysical observations.

In this paper, we extend our previous work \cite{DAgostino:2025yej} by investigating gravitational perturbations of the DD BH. At the same time, we generalize the results of \cite{Chen:2021pxd} to the case of a generic static and spherically symmetric spacetime.
In particular, we analytically study the propagation of both axial and polar perturbations 
and search for a wave-equation formulation that includes additional contributions sourced by the nonlocal scalar fields. This framework allows us to describe the wave functions and the corresponding potentials as first-order corrections to their Schwarzschild counterparts. 

The structure of the paper is as follows. We review in Sec.~\ref{sec:DW-nonlocal} the main features of the DD BH spacetime solution. After discussing the most general form of a time-dependent, axisymmetric metric, Sec.~\ref{sec:gravitational-perturbations} is devoted to the analysis of axial and polar perturbations around the DD BH metric. 
Finally, in Sec.~\ref{sec:conclusions}, we summarize the main results of our study and discuss our conclusions and potential directions for future works.

Throughout this work, we use geometric units, namely $\hbar=G=c=1$, and set the BH mass to unity. The flat metric is indicated by $\eta_{\mu\nu}=\rm{diag}(-1,1,1,1)$. 

\section{Nonlocal DD black hole}
\label{sec:DW-nonlocal}

The DD BH solution arises within the framework of the revised DW nonlocal gravity action \cite{Deser:2019lmm}
\begin{equation} \label{eq:nonlocal-action}
S=\dfrac{1}{16\pi}\int \sqrt{-g}\, R\left[1+f(Y)\right]\dd^4 x \,,
\end{equation}
where $f(Y)$ is the distortion function encoding the nonlocal effects, $R$ is the Ricci scalar, and $g$ denotes the determinant of the metric tensor, $g_{\mu\nu}$.

The dynamics of this model can be studied through a localized form involving four independent auxiliary scalar fields governed by the following equations of motion:
\begin{subequations} \label{eq:dyn-SF}
\begin{align}
    \Box X&=R\,, \label{eq:X} \\ 
    \Box Y&=g^{\mu\nu}\partial_\mu X\partial_\nu X\,, \label{eq:Y}\\
    \Box U&=-2\nabla_\mu (V\nabla^\mu X)\,, \label{eq:U} \\
    \Box V&=R f_{,Y}\,,\label{eq:V}
\end{align}
\end{subequations}
where $\Box\equiv \nabla_\mu \nabla^\mu$ is the d'Alembert operator such that
\begin{equation} 
    \Box u\equiv\frac{1}{\sqrt{-g}}\partial_\alpha\left[\sqrt{-g}\,\partial^\alpha u\right],
    \label{eq:box-operator}
\end{equation}
with $u$ being a generic scalar field.

The vacuum field equations are obtained by varying the action \eqref{eq:box-operator} with respect to $g_{\mu\nu}$, leading to \cite{Deser:2019lmm}
\begin{align}
\left(G_{\mu\nu}+g_{\mu\nu}\Box-\nabla_\mu \nabla_\nu\right)W+\mathcal{K}_{(\mu\nu)}-\frac{1}{2}g_{\mu\nu}g^{\alpha\beta}\mathcal{K}_{\alpha\beta} =0\,,
\label{eq:FE}
\end{align}
where $G_{\mu\nu}= R_{\mu\nu}-\frac{1}{2}g_{\mu\nu}R$ is the Einstein tensor, and $W\equiv 1+U+f(Y)$. Moreover,
\begin{equation}
\mathcal{K}_{\mu\nu}\equiv \partial_\mu X\partial_\nu U +\partial_\mu Y \partial_\nu V+V\partial_\mu X \partial_\nu X \,,
\label{eq:Kmunu}
\end{equation}
and $\mathcal{K}_{(\mu\nu)}\equiv (\mathcal{K}_{\mu\nu}+\mathcal{K}_{\nu\mu})/2$.

We thus look for a static and spherically symmetric solution of Eq.~\eqref{eq:FE}, in spherical-like coordinates:
\begin{equation}
\dd s^2=g_{tt}(r) \dd t^2+g_{rr}(r)\dd r^2 +r^2(\dd \theta^2+\sin^2\theta\, \dd\varphi^2)\,.
    \label{eq:general-metric}
\end{equation}
For this aim, we relate the tetrad field of a static observer at infinity, $\left\{{\bf e}_t\,, {\bf e}_r\,,{\bf e}_\theta\,,{\bf e}_\varphi\right\}$, to that linked to a static observer in the spacetime \eqref{eq:general-metric}, namely \cite{Morris:1988cz}
\begin{align}
{\bf e}_{\hat t} = \frac{{\bf e}_t}{\sqrt{-g_{tt}}}\,,\ 
{\bf e}_{\hat r}= \frac{{\bf e}_r}{\sqrt{g_{rr}}}\,,\
{\bf e}_{\hat \theta} =\frac{{\bf e}_\theta}{r}\,,\ 
{\bf e}_{\hat \varphi}=\frac{{\bf e}_\varphi}{r\sin\theta}\,,
\label{eq:proper_tetrad}
\end{align}
such that $g_{\hat\mu\hat\nu}=e_{\hat \mu}^\alpha e_{\hat \nu}^\beta g_{\alpha\beta}\equiv\eta_{\mu\nu}$, where
\begin{subequations} \label{eq:tetrad}
\begin{align}
    e^{\hat a}_{\mu}&=\text{diag}\left(\sqrt{-g_{tt}},\sqrt{g_{rr}},r,r\sin\theta\right),\label{eq:general-tetrad-inverse} \\
    e_{\hat a}^{\mu}&=\text{diag}\left(\frac{1}{\sqrt{-g_{tt}}},\frac{1}{\sqrt{g_{rr}}},\frac{1}{r},\frac{1}{r\sin\theta}\right).
    \label{eq:general-tetrad}
    \end{align}    
\end{subequations}
Any vector $u_a$ or tensorial quantity $T_{\alpha\beta}$ is transformed in the tetrad frame via Eq.~\eqref{eq:general-tetrad-inverse} as $u_{\hat a}=e_{\hat a}^\mu u_\mu$ and $T_{\hat{a}\hat{b}}=e^\mu_{\hat a}e^\nu_{\hat b} T_{\mu\nu}$, respectively.

Therefore, we recast Eq.~\eqref{eq:FE} in the tetrad frame as \cite{Chen:2021pxd}
\begin{align}
&R_{\hat a\hat b}W-e_{\hat a}^\mu(W_{,\hat b})_{,\mu}+\gamma_{\hat c\hat b \hat a}W_{,\hat d}\eta^{cd}\nonumber\\
&+\eta_{ab}\left(\Box W-\frac{R}{2}W-\frac{1}{2}\mathcal{K}^\rho_\rho\right)-e^\rho_{(\hat a}e^\lambda_{\hat b)}\mathcal{K}_{\rho\lambda}=0\,,
\label{eq:linearized}
\end{align}
where $\gamma_{\hat c\hat a\hat b}\equiv e_{\hat b}^{\mu}\nabla_\mu e_{\hat c}^{\nu} e_{\hat a\nu}$.
In so doing, we obtain the DD BH solution (see Ref. \cite{DAgostino:2025wgl} for more details): 
\begin{subequations}\label{eq:BH_sol}
\begin{align}
    -g_{tt}&=1-\frac{2}{r}-\frac{\alpha }{r^k}\,, \label{eq:BH_solA}\\
    (g_{rr})^{-1}&=1-\frac{2}{r}+\frac{\alpha}{3^{k} r^{k+1}(r-3)^2} \left\{3^k r \Big{[}k (r-3) (r-2)\right.\notag\\
    &\left.\quad+4 r-9\Big{]}-3 (r-2) (2 r-3) r^k\right\}.
    \label{eq:BH_solB}
\end{align}    
\end{subequations}
These expressions define a family of solutions parametrized by the real constant $k>1$. The DD BH constitutes a first-order deviation from the Schwarzschild metric, characterized by the small parameter $0<\alpha \ll 1$. 
Accordingly, all physical quantities shall be evaluated to first-order in $\alpha$. These solutions satisfy the condition of asymptotic flatness and are free from essential singularities (except in $r=0$) outside the event horizon.  
In particular, the Ricci scalar is given by\footnote{We note that the scalar curvature is well defined in $r=3$:
\begin{equation*}
    \lim_{r\to3}R(r)=\frac{\alpha k\left(5+6k-k^2\right)}{3^{k+3}}\,.
\end{equation*}}
\begin{align}
    &R(r)=\alpha\frac{3r^{-k-\frac{5}{2}}}{(r-3)^3} \bigg[3 \left(7 k^2-13 k+4\right) r^{\frac{3}{2}}+k(k-1) r^{\frac{7}{2}}\notag\\
    &+4 k (3-2 k) r^{\frac{5}{2}} -\frac{4r^{k+\frac{3}{2}}}{3^{k-1}}+\frac{2r^{k+\frac{1}{2}}}{3^{k-2}} -18 (k-1)^2 r^{\frac{1}{2}}\bigg].
    \label{eq:Ricci}
\end{align}
The analytical expressions of the auxiliary scalar fields corresponding to the DD BH solution are \cite{DAgostino:2025wgl}
\begin{subequations} \label{eq:SCALAR-FIELDS}
\begin{align}
X(r)&=\alpha  \left[\frac{3^{2-k}-3 r^{1-k}}{3-r}+x_1 \ln \left(\frac{r}{r-2}\right)\right],\label{eq:X_sol}\\
Y(r)&=\alpha  \ln \left(\frac{r}{r-2}\right),
\label{eq:Ysol1}\\
V(r)&=\alpha  \left[\frac{\mathcal{P}_k(r)}{r^{2k-2}}+v_1 \ln \left(\frac{r}{r-2}\right)\right],
    \label{eq:V_sol}\\
U(r)&=0\label{eq:U_sol}\,, \\
W(r)&=1+\frac{\alpha  \left(3^{1-k}-r^{1-k}\right)}{3-r}\,,\label{eq:W_sol}
\end{align}    
\end{subequations}
where $x_1$ and $v_1$ are arbitrary real constants, and $\mathcal{P}_k(r)$ is a polynomial of order $2k-3$ in the radial coordinate $r$. 
In this case, the distortion function takes the form 
\begin{equation}\label{eq:fY}
f(Y)=\frac{\alpha  \left(e^{Y/\alpha }-1\right)}{e^{Y/\alpha }-3}  \left[3^{1-k}-\left(\frac{2}{e^{Y/\alpha }-1}+2\right)^{1-k}\right],
\end{equation}
whose limit $f(Y)\to 0$ reproduces GR.

It is worth remarking that, within the domain of the DD perturbative solution, the future spacelike singularity at $r=0$ persists and is not qualitatively altered. Although the corrections become increasingly important in the interior region and modify the detailed behavior of curvature invariants, they do not remove or regularize the singularity, nor do they change its causal character. In particular, the singularity remains spacelike and constitutes the future endpoint of timelike and null geodesics inside the event horizon.

\section{Gravitational perturbations}
\label{sec:gravitational-perturbations}

This section is devoted to study the gravitational perturbations of the DD BH. We begin by considering time-dependent axisymmetric modes, for which a suitable form of the metric is given by \cite{1972ApJ...175..379C}
\begin{align}
    \dd s^2=&-e^{2\nu}\dd t^2+e^{2\mu_2}\dd r^2+e^{2\mu_3}\dd\theta^2\notag \\
    &+e^{2\psi}\left(\dd\varphi-\sigma \dd t-q_2\dd r-q_3 \dd\theta\right)^2,
    \label{metric:axisymmetric}
\end{align}
where $\nu,\,\mu_2,\,\mu_3\,\psi,\,\sigma,\,q_2,$ and $q_3$ are functions of $(t,r,\theta)$.

Gravitational perturbations decompose into axial and polar  sectors, which are generally treated separately. The former induce a rotation to the BH, whereas the latter do not. As a consequence, axial modes generate the non-vanishing perturbations $\sigma$, $q_2$, and $q_3$, while $\nu,\mu_2,\mu_3$ can be fixed at their background values (see Sec.~\ref{sec:axial}). Instead, polar perturbations are characterized by non-zero small increments of $\nu$, $\mu_2$, $\mu_3$, and $\psi$ around the metric~\eqref{metric:axisymmetric}, while we can set $q_2=q_3=\sigma=0$ (see Sec.~\ref{sec:polar}). 

In the following, we keep our treatment as general as possible so as to include any static and spherical symmetric spacetime arising from the revised DW nonlocal theory. Towards the concluding parts, we shall specialize to the case of the DD BH to present the final results. For our purposes, we follow the strategy outlined in Ref.~\cite{1972ApJ...175..379C}, while generalizing the results obtained in Ref.~\cite{Chen:2021pxd}.

\subsection{Axial perturbations}
\label{sec:axial}

For axial modes, the metric~\eqref{metric:axisymmetric} takes the following form at first-order perturbations:
\begin{equation}
g_{\mu\nu}=
\begin{pmatrix}
 -e^{2 \nu} & 0 & 0 &-e^{2 \psi} \sigma \\
 0 & e^{2 \mu_2} & 0 &   -e^{2 \psi} q_2 \\
 0 & 0 & e^{2 \mu_3} &   -e^{2 \psi} q_3 \\
   -e^{2 \psi} \sigma &   -e^{2 \psi} q_2 &   -e^{2 \psi} q_3 & e^{2 \psi} \\
\end{pmatrix}.
\end{equation}
In this context, we consider the $(\varphi,r)$ and $(\varphi,\theta)$ components of the linearized field equations~\eqref{eq:linearized}, whose expressions are given by, respectively,
\begin{subequations}
\begin{align}
   &\left(\sigma_{,r}-q_{2,t}\right)_{,t}+2rq_2\bigg\{W\left(1-e^{2\mu_2}+r\nu_{,r}-r\mu_{2,r}\right) \sin\theta\notag \\
   &+e^{\mu_2}\Big[W\left(2\mu_{2,r}-r\nu_{,r}^2+r\nu_{,r}\mu_{2,r}-r\nu_{,rr}\right)-rW_{,rr}\notag \\
   &+W_{,r}\left(1+r\mu_{2,r}\right)\Big]\bigg\}+\frac{e^{\nu+\mu_2}}{r^4\sin^3\theta}Q_{,\theta}=0\,, \label{eq:axial_12}
\\
    &2e^{2\nu}q_3(1-\sin\theta)\left[rW_{,r}+W(1-e^{2\mu_2}+r\nu_{,r}-r\mu_{2,r})\right]\notag \\
    &+2e^{2\nu}q_2\left(r\sin\theta-e^{2\mu_2}\right)W_{,r}+\frac{e^{\nu+\mu_2}}{\sin^2\theta}W_{,r}Q\notag \\
    &+\frac{e^{\nu-\mu_2}}{r^2\sin^3\theta}Q_{,r}-\left(\sigma_{,\theta}-q_{3,t}\right)_{,t}=0\,,
    \label{eq:axial_13}
\end{align} 
\end{subequations}
where we have defined
\begin{equation} \label{eq:Q}
    Q(t,r,\theta)\equiv e^{\nu-\mu_2}\,r^2\sin^3 \theta\, Q_{23}(t,r,\theta)\,,
\end{equation}
with $Q_{AB}\equiv q_{A,B}-q_{B,A}$ for $A,B=(2,3)$.

Then, under the decompositions
\begin{subequations}
\begin{align}
&q_2(t,r,\theta)=\tilde q_2(r,\theta)e^{-i\omega t}\,,\\
&q_3(t,r,\theta)=\tilde q_3(r,\theta)e^{-i\omega t}\,,\\
&Q(t,r,\theta)=\tilde Q(r,\theta)e^{-i\omega t}\,,
\end{align}
\end{subequations}
from Eqs.~\eqref{eq:axial_12} and \eqref{eq:axial_13} we obtain, respectively,
\begin{align}
&\frac{ e^{\nu+\mu_{2}} }{r^4\sin^3\theta } \tilde Q_{,\theta}- i\omega\tilde\sigma_{,r}+\frac{\tilde q_{2}}{r^2}\Big\{2 e^{2(\nu-\mu_2)} \left(1+r \nu_{,r}-r \mu_{2,r}\right)\notag\\
&-\frac{2 e^{2 \nu-\mu_2}}{W\sin\theta} \left[W \left(r\nu_{,rr}-r\nu_{,r} \mu_{2,r}+r\nu_{,r}^2-2 \mu_{2,r}\right)+r W_{,rr}\right.\notag\\
&\left.-\left(1+r \mu_{2,r}\right)-W_{,r}\right]-2 e^{2 \nu}+r^2 \omega ^2\Big\}=0\,,
\end{align}
and
\begin{align}
&\frac{ e^{\nu-\mu_{2}} }{r^2\sin^3\theta}\left( \tilde Q_{,r}+\tilde{Q}\frac{W_{,r}}{W}\right)+ i\omega \tilde\sigma_{,\theta}- \tilde q_{3}\omega^{2}\notag\\
&- \tilde q_{3}\bigg\{\frac{2 (1-\sin\theta) e^{2(\nu-\mu_2)}}{r^2 W\sin\theta} \Big[W (e^{2\mu_2}-r \nu_{,r}+r \mu_{2,r}-1)\notag \\
&-r W_{,r}\Big]\bigg\}-\frac{2 \tilde q_2 e^{2(\nu-\mu_2)}W_{,r}\left( e^{\mu_2}-r\sin\theta\right)}{r W\tan\theta\sin\theta}=0\,.
\end{align}
By differentiating Eq.~\eqref{eq:axial_12} with respect to $\theta$ and Eq.~\eqref{eq:axial_13} with respect to $r$, and then combining the results, one can eliminate $\tilde{\sigma}_{,r\theta}$. This yields
\begin{align}
  &\tilde{Q}\bigg\{W^2 \Big[2 e^{2 \nu} \left(e^{2 \mu_2}-r \nu_{,r}+r \mu_{2,r}-1\right)-r^2 \omega ^2 e^{2 \mu_2}\Big]\notag\\
  &-r e^{2 \nu} W \Big[W_{,r} \left(r \nu_{,r}-r \mu_{2,r}-2\right)+r W_{,rr}\Big]+e^{2 \nu} r^2  W_{,rr}^2 
  \bigg\}\notag \\
  &-e^{2\nu}W\bigg\{r^2W_{,r}\tilde{Q}_{,r}+e^{2\mu_2}W(\tilde{Q}_{,\theta\theta}-3Q_{,\theta}\cot\theta)\notag \\
  &+rW\left[r\tilde{Q}_{,rr}-(2-r\nu_{,r}+r\mu_{,r})\tilde{Q}_{,r}\right]\!\bigg\}= 0\,.
  \label{eq:axial_comb}
\end{align}

It is worth remarking that an approximation is needed in order to obtain Eq.~\eqref{eq:axial_comb}. In fact, when the DD spacetime is considered, the terms depending on the functions $q_2$, $q_3$, $q_{2,r}$, and $q_{3,r}$ appear only in products of the forms $\sim\alpha\, q_{i}$ and $\sim\alpha\, q_{i,r}$. These represent very small quantities since $\alpha\ll 1$ and $q_i,q_{i,r}\ll1$, so we have neglected them for mathematical convenience.

Now, we separate the variables as 
\begin{equation}\label{eq:spli-Q}
    \tilde{Q}(r,\theta)=\mathbb{Q}(r)C_{q}^{(m)}(\theta)\,,
\end{equation}
where $C_{q}^{(m)}$ is the Gegenbauer function, satisfying the  differential equation
\begin{equation}\label{eq:ODE-G}
    \left[\frac{d}{d\theta}(\sin^{2m}\theta) \dfrac{d}{d\theta}+q(q+2m)\sin^{2m}\theta\right]C_q^{(m)}(\theta)=0\,.
\end{equation}
In particular, for $q=l+2$ and $m=-3/2$, as in the Schwarzschild case \cite{1972ApJ...175..379C}, one has 
\begin{equation} \label{eq:Geg-func}
C_{l+1}^{-3/2}(\theta)=\Big[P_{l,\theta\theta}(\cos\theta)-P_{l,\theta}(\cos\theta)\cot\theta\Big]\sin^2\theta,
\end{equation}
where $P_l(\cos\theta)$ is the Legendre polynomial satisfying 
\begin{equation}
    P_{l,\theta\theta}(\cos\theta)+\cot\theta\, P_{l,\theta}(\cos\theta)+l(l+1)P_l(\cos\theta)=0\,.
    \label{eq:Legendre}
\end{equation} 
Using Eq.~\eqref{eq:Geg-func} in Eq.~\eqref{eq:spli-Q}, and exploiting Eq.~\eqref{eq:Legendre}, Eq.~\eqref{eq:axial_comb} reduces to the radial equation
\begin{align}
&\mathbb{Q}_{,rr}+\left(\nu_{,r}-\mu_{2,r}-\frac{2}{r}+\frac{W_{,r}}{W}\right)\mathbb{Q}_{,r}\notag\\
&+\left\{\frac{e^{2\mu_2}}{r^2} \left[e^{-2 \nu } r^2 \omega ^2-2 (n+1)\right]+\frac{2}{r^2}(1+r\nu_{,r}-r\mu_{2,r})\right.\notag\\
&\left.+\frac{W_{,rr}}{W}+\frac{W_{,r}}{W}\left(\nu_{,r}-\mu_{,r}-\frac{2}{r}-\frac{1}{W}\right)\right\}\mathbb{Q}\simeq 0\,,
\label{eq:AXIAL}
\end{align}
where $2n\equiv(l+2)(l-1)$. We notice that the separation of the radial and the angular parts is possible because we have neglected the terms accompanied by $\csc\theta$, which scale as $\sim \alpha/r^6$ for the DD BH and are subdominant compared to the other contributions.

It is straightforward to verify that, taking the DD BH in the limit $\alpha\to 0$, Eq.~\eqref{eq:AXIAL} leads to the standard Regge--Wheeler equation \cite{Regge:1957td}. 
However, for $\alpha\neq 0$, the radial equation cannot be recast into a Schr\"odinger-like form. Indeed, introducing the tortoise radius $r_*$, it is possible to evaluate the Jacobian of the transformation as \cite{DAgostino:2025yej} 
\begin{align}
    \frac{\dd r}{\dd r_*}&=1-\frac{2}{r}+\alpha\bigg\{\frac{3^{-k}r^{1-k}}{2(r-3)^2}\Big[(9-6 r) (r-2) r^k\notag \\
    &+3^k r \big(k (r-3) (r-2)-(r-10) r-18\big)\Big]\bigg\}\,.
    \label{eq:tortoise}
\end{align}
Therefore, from  Eq.~\eqref{eq:AXIAL}, we finally find
\begin{align}
    \Psi_{,r_*r_*}^{(\text{ax})}+\big[\omega^2-V^\text{(ax)}\big]\Psi^{(\text{ax})}=\beta_1\mathbb{Q}_{,r}+\beta_2\mathbb{Q}_{,rr}\,,
    \label{eq:axial_wave}
\end{align}
where $\Psi^{(\text{ax})}=\Psi_0^{\text{ax}}+\alpha \Psi_1^{\text{ax}}$, with 
\begin{subequations}
\begin{align}
\Psi_0^{(\text{ax})}=&\ \frac{\mathbb{Q}}{r}\,, \\
\Psi_1^{\text{ax}}=&\ \frac{3^{-k} r^{3-k}\mathbb{Q}}{\left(r^2-5 r+6\right)^2} \bigg[18 r^k-9 r^{k+1}+3^k (k-2) r^3\notag\\
&-3^k (5 k-14) r^2+3^{k+1} (2 k-7) r\bigg]\,.
\end{align}
\end{subequations}
Also, the effective potential takes the form $V^{(\text{ax)}}=V_0^{(\text{ax})}+\alpha V_1^{(\text{ax})}$. Specifically, we have
\begin{align}
    &V_0^{(\text{ax})}= \frac{2(r-2)}{r^4}\big[(n+1)r-3\big]\,,
\end{align}
which coincides with the Regge-Wheeler potential \cite{Regge:1957td,Chandrasekhar:1984siy}. The first-order correction $V_1^{\text{(ax)}}$ is reported in Appendix~\ref{app:axial} (see Eq.~\eqref{eq:V1_axial}), together with the  expressions of $\beta_1$ and $\beta_2$ (see Eqs.~\eqref{eq:beta1} and \eqref{eq:beta2}).

It is important to stress that the nonlocal gravitational effects alter the standard wave equation due to the presence of non-zero terms in the right-hand side of Eq.~\eqref{eq:axial_wave}. The standard GR outcomes can, in fact, be recovered only when $W=\text{const}$. This result is consistent with that previously obtained in Ref.~\cite{Chen:2021pxd}. Conversely, in the most general case where $W_{,r}\neq0$, one must resort to numerical methods to fully determine the propagation of the axial modes.

\subsection{Polar perturbations}
\label{sec:polar}

In the case of polar modes, the non-zero components of the metric tensor at the first-order perturbations read 
\begin{subequations} \label{eq:linear-pert-met}
\begin{align}
    &g_{tt}=-e^{2\bar\nu}\left(1+2\delta\nu\right),\quad g_{rr}= e^{-2\bar\mu_2}\left(1+2\delta\mu_2\right),\\
    &g_{\theta\theta}=e^{2\bar\mu_3}\left(1+2\delta\mu_3\right),\quad g_{\phi\phi}=e^{2\bar\psi}\left(1+2\delta\psi\right),
\end{align}
\end{subequations}
with $\delta\nu,\delta\psi,\delta\mu_2,\delta\mu_3\ll 1$. In particular, we consider the following small-increment decompositions:
\begin{subequations} \label{eq:decomp-metric}
\begin{align}
\nu(t,r,\theta)&=\bar\nu(r)+\delta\nu(r,\theta)e^{-i\omega t}, \\   
\psi(t,r,\theta)&=\ln(r\sin\theta)+\delta\psi(r,\theta)e^{-i\omega t}, \\   
\mu_2(t,r,\theta)&=\bar\mu_2(r)+\delta\mu_2(r,\theta)e^{-i\omega t}, \\  
\mu_3(t,r,\theta)&=\ln r+\delta\mu_3(r,\theta)e^{-i\omega t},   
\end{align}
\end{subequations}
where the overbar quantities mean that they are evaluated at the DD BH background values \eqref{eq:BH_sol}. 

The scalar fields follow a similar decomposition:
\begin{subequations}\label{eq:pert-SF}
\begin{align}
X(t,r,\theta)&=\bar X(r)+\delta X(r,\theta)e^{-i\omega t},\\
Y(t,r,\theta)&=\bar Y(r)+\delta Y(r,\theta)e^{-i\omega t},\\
U(t,r,\theta)&=\bar U(r)+\delta U(r,\theta)e^{-i\omega t},\\
V(t,r,\theta)&=\bar V(r)+\delta V(r,\theta)e^{-i\omega t},\\
W(t,r,\theta)&=\bar W(r)+\delta W(r,\theta)e^{-i\omega t},
\end{align}    
\end{subequations}
where the following relation can be obtained from the background scalar fields \eqref{eq:SCALAR-FIELDS}:
\begin{equation}
\bar U_{,r}\bar X_{,r}+\bar V_{,r}\bar Y_{,r} +\bar V_{,r} (\bar X_{,r})^2=0\,. 
\label{eq:rel_salar}
\end{equation}
The above perturbations can be substituted into Eq.~\eqref{eq:linearized} to obtain the non-vanishing linearized components of Eq.~\eqref{eq:FE}. In doing so, we follow the approach of Ref.~\cite{Chandrasekhar:1984siy}. A similar strategy was adopted in Ref.~\cite{Chen:2021pxd} within the revised DW framework, where the linear perturbations of the field equations were derived by assuming the Schwarzschild metric as the background spacetime. In contrast, in the present study, the perturbations are calculated around the DD BH \cite{DAgostino:2025wgl}. We also note that some caution is required when using the results of Ref.~\cite{Chen:2021pxd} due to some  inaccuracies in the reported components of the linearized field equations. These have now been corrected in Appendix~\ref{sec:Sch-FE}.

In our case, we find that the $(t,r)$, $(t,\theta)$, and $(r,\theta)$ components are given by, respectively,
\begin{subequations}
\begin{align}
&\bar W\left[\left(\delta\psi+\delta\mu_3\right)_{,r}+\left(\frac{1}{r}-\bar\nu_{,r}\right)\left(\delta\psi+\delta\mu_3\right)-\frac{2}{r}\delta\mu_2\right]\nonumber\\
&-\bar W_{,r}\delta\mu_2+e^{\bar\nu}\left(e^{-\bar\nu}\delta W\right)_{,r}
=\frac{1}{2}\big(\bar X_{,r}\delta U+\bar U_{,r}\delta X \nonumber\\
&+\bar V_{,r}\delta Y+\bar Y_{,r}\delta V+2\bar V\bar X_{,r}\delta X\big)\,,
\label{eq:02}\\
&\bar W\!\left[\left(\delta\psi+\delta\mu_2\right)_{,\theta}+\left(\delta\psi-\delta\mu_3\right)\cot\theta\right]+\delta W_{,\theta}=0\,,
\label{eq:03}\\
&\bar W\!\left[\left(\delta\nu+\delta\psi\right)_{,r\theta}+\left(\delta\psi-\delta\mu_3\right)_{,r}\cot\theta-\!\left(\frac{1}{r}-\bar\nu_{,r}\right)\delta\nu_{,\theta}\right.\nonumber\\
&\left.-\left(\frac{1}{r}+\bar\nu_{,r}\right)\delta\mu_{2,\theta}\right]+\delta W_{,r\theta}-\frac{1}{r}\delta W_{,\theta}-\bar W_{,r}\delta\mu_{2,\theta}  \nonumber\\
&=\frac{1}{2}\big(\bar X_{,r}\delta U_{,\theta}+\bar U_{,r}\delta X_{,\theta}+\bar V_{,r}\delta Y_{,\theta}+\bar Y_{,r}\delta V_{,\theta}\notag \\
&+2\bar V\bar X_{,r}\delta X_{,\theta}\big).
\label{eq:23}
\end{align}
\end{subequations}
Additionally, the $(r,r)$  component is
\begin{align}
    &e^{2\bar\mu_2}\!\left\{\!\frac{2}{r}\left[\delta\nu_{,r}\!-\left(\frac{1}{r}+2\bar\nu_{,r}\!\right)\!\delta\mu_2\right]+\!\left(\frac{1}{r}+\bar\nu_{,r}\!\right)\left(\delta\psi+\delta\mu_3\right)_{,r}\!\right\}\notag \\
    &+\frac{1}{r^2}\left[2\delta\mu_3+\left(\delta\nu+\delta\psi\right)_{,\theta\theta}+\left(2\delta\psi+\delta\nu-\delta\mu_3\right)_{,\theta}\cot\theta\right]\notag \\
    &+e^{-2\bar\nu}\omega^2(\delta\psi+\delta\mu_3)
    =\frac{-1}{\bar W}\bigg\{e^{2\bar\mu_2}\bigg[\left(\frac{1}{r}+2\bar\nu_{,r}\right)\frac{\delta W}{r}\notag \\
    &+\!\left(\frac{2}{r}+\bar\nu_{,r}\!\right)\!\delta W_{,r}+\bar W_{,r}\!\left(\!\left(\delta\psi+\delta\mu_3+\delta\nu\right)_{,r}\right.\notag\\
    &\left. -2\delta\mu_2\Big(\frac{2}{r}+\bar\nu_{,r}\Big)\!\right)\!\bigg]+e^{-2\bar\nu}\omega^2\delta W+\frac{1}{r^2}\left(\delta W_{\theta\theta}-\delta W\right.\notag\\
    &\left.+\delta W_{,\theta}\cot\theta\right)-\!\left(\frac{1}{2}-e^{2\bar\mu_2}\!\right)\!\left(\bar X_{,r}\delta U_{,r}+U_{,r}\delta X_{,r}+\bar V_{,r}\delta Y_{,r}\right.\nonumber\\
    &\left.+\bar Y_{,r}\delta V_{,r}+\bar X_{,r}^2\delta V+2\bar V\bar X_{,r}\delta X_{,r}\right)\!\bigg\}\,,\label{eq:22}
\end{align}
while the $(\varphi,\varphi)$ component is given by
\begin{align}
    &e^{2\bar\mu_2}\bigg[\left(\delta\nu+\delta\mu_3\right)_{,rr}-\frac{2\delta\mu_2}{r}\left(\bar\nu_{,r}+r\bar\nu_{,r}^2+r\bar\nu_{,rr}\right)\notag\\
    &-\left(\frac{1}{r}+\bar\nu_{,r}\right)\delta\mu_{2,r}+\left(\frac{2}{r}+\bar\nu_{,r}\right)\delta\mu_{3,r}+\left(\frac{1}{r}+2\bar\nu_{,r}\right)\delta\nu_{,r}\notag\\
    &+\bar\mu_{2,r}\!\left(\!\delta\nu_{,r}+\delta\mu_{3,r}-2\delta\mu_2\Big(\frac{1}{r}+\bar\nu_{,r}\Big)\right)\!\bigg] +\frac{1}{r^2}\left(\delta\nu+\delta\mu_2\right)_{,\theta\theta}\notag \\
    &=\frac{-1}{\bar W}\bigg\{e^{-2\bar\nu}\omega^2\left[\delta W+\bar W(\delta\mu_2+\delta\mu_3)\right]+\frac{\delta W_{,\theta\theta}}{r^2}\notag \\
    & +e^{2\bar\mu}\bigg[\bar\nu_{,rr}\delta W +\left(\frac{1}{r}+\bar\nu_{,r}\right)(\bar\nu_{,r}\delta W +\delta W_{,r})+\delta W_{,rr}\notag \\
    &+\bar\mu_{2,r}\left(\delta W\Big(\frac{1}{r}+\bar\nu_{,r}\Big)+\delta W_{,r}\right)-2\delta\mu_2 \bar W_{,rr}\notag \\
    &+ \bar W_{,r}\left(\left(\delta\nu+\delta\mu_3-\delta\mu_2\right)_{,r}-2\delta\mu_2\Big(\frac{1}{r}+\bar\nu_{,r}+\bar\mu_{2,r}\Big)\right)\!\bigg]\notag \\
    &-\frac{1}{2}(\bar X_{,r}\delta U_{,r}+\bar U_{,r}\delta X_{,r}+\bar V_{,r}\delta Y_{,r}+\bar Y_{,r}\delta V_{,r}\nonumber\\
    &+\bar X_{,r}^2\delta V+2\bar V\bar X_{,r}\delta X_{,r})\!\bigg\}\,.\label{eq:44}
\end{align}

To simplify the above equations, we apply the following decompositions into polar and radial parts \cite{Chandrasekhar:1984siy}:
\begin{subequations}
\begin{align}
\delta\nu&=\mathbb{N}(r)P_l(\cos\theta)\,,\\
\delta\mu_2&=\mathbb{L}(r)P_l(\cos\theta)\,,\\
\delta\mu_3&=\mathbb{T}(r)P_l(\cos\theta)+\mathbb{V}(r)P_{l,\theta\theta}(\cos\theta)\,,\\
\delta\psi&=
\mathbb{T}(r)P_l(\cos\theta)+\mathbb{V}(r)P_{l,\theta}(\cos\theta)\cot\theta\,,\\
\delta W&=\bar W(r)\delta\tilde{W}(r)P_l(\cos\theta)\,.
\end{align}
\label{ansatz}
\end{subequations}
Additionally, we consider
\begin{align}
\mathbb{U}&=\frac{\tilde{\mathbb{U}}(r)}{r}P_l(\cos\theta)\,\nonumber\\
&\equiv \frac{\bar X_{,r}\,\delta U+ (\bar U_{,r}+2\bar V\bar X_{,r})\delta X +\bar V_{,r}\,\delta Y +\bar  Y_{,r}\,\delta V}{2\bar W}\,.
\label{U}
\end{align}
Making use of Eqs.~\eqref{eq:Legendre}, \eqref{ansatz}, and \eqref{U}, we can recast Eqs.~\eqref{eq:02}, \eqref{eq:03}, and \eqref{eq:23} as follows:
\begin{subequations}
\begin{align}
   & \left[\frac{\dd}{\dd r}+\left(\frac{1}{r}-{\bar\nu_{,r}}\right)\right]\left[2\mathbb{T}-l(l+1)\mathbb{V}\right]-\mathbb{L}\left(\frac{2}{r}+\frac{\bar W_{,r}}{\bar W}\right) \notag \\
   &+\delta\tilde W\frac{\bar W_{,r}}{\bar W} +e^{\bar\nu}\left(e^{-\bar\nu}\delta\tilde{W}\right)_{,r}=\frac{\tilde{\mathbb{U}}}{r}\,,
   \label{eq:02new}\\
&\mathbb{L}+\mathbb{T}+\delta\tilde{W}=\mathbb{V}\,,
\label{eq:03new}\\
&(\mathbb{N}+\mathbb{T}-\mathbb{V})_{,r}-\mathbb{N}\left(\frac{1}{r}-\bar\nu_{,r}\right)-\mathbb{L}\left(\frac{1}{r}+\bar\nu_{,r}+\frac{\bar W_{,r}}{\bar  W}\right) \notag \\
    &+\delta\tilde W\frac{ \bar W_{,r}}{\bar  W}-\frac{1}{r}\delta\tilde{W}+\delta\tilde{W}_{,r}=\frac{\tilde{\mathbb{U}}}{r} \,. 
    \label{eq:23new}
\end{align}
\end{subequations}
Moreover, Eq.~\eqref{eq:22} becomes
\begin{align}
   & e^{2\bar\mu_2}\bigg\{\mathbb{N}_{,r}\!\left(\!\frac{2}{r}+\frac{\bar W_{,r}}{\bar W}\!\right)\!+\!\left(\!\frac{1}{r}+\bar\nu_{,r}+\frac{\bar W_{,r}}{\bar W}\!\right)\!\left[2\mathbb{T}-l(l+1)\mathbb{V}\right]_{,r}\notag \\
    &-(2\mathbb{L}-\delta \tilde W)\left[\frac{1}{r}\left(\frac{1}{r}+2\bar\nu_{,r}\right)+\frac{\bar W_{,r}}{\bar W}\left(\frac{2}{r}+\bar\nu_{,r}\right)\right]\notag  \\
    &+\left(\frac{2}{r}+\bar\nu_{,r}\right)\delta \tilde W_{,r}\bigg\}-\frac{1}{r^2}\left[l(l+1)\mathbb{N}+(l+2)(l-1)\mathbb{T})\right]\notag \\
    &-\frac{\delta\tilde W}{r^2}\left[l(l+1)+1\right]+e^{-2\bar \nu}\omega^2\left[2\mathbb{T}-l(l+1)\mathbb{V}+\delta\tilde{W}\right]\notag \\
    &=\left(1-2e^{2\bar\mu_2}\right)\left(\frac{\tilde{\mathbb{U}}}{r}\right)_{,r} ,
    \label{eq:22new}
\end{align}
where we have used the following approximation:
\begin{align}
    &\bar X_{,r}\delta U_{,r}+\bar U_{,r}\delta X_{,r}+\bar V_{,r}\delta Y_{,r}+\bar Y_{,r}\delta V_{,r}+\bar X_{,r}^2\delta V \notag \\
    &+2\bar V\bar X_{,r}\delta X_{,r}\simeq 2P_l\bar W\left(\frac{\tilde{\mathbb{U}}}{r}\right)_{,r}\,,
    \label{eq:rel_scalar_2}
\end{align}
which holds true if we neglect the second-order derivatives of the auxiliary fields at the background\footnote{In view of Eqs.~\eqref{eq:X_sol}--\eqref{eq:U_sol}, we have $\bar U_{,r}\bar X_{,r}=\bar U_{,rr}=0$, and $\bar X_{,r}^2=\bar V_{,r}\bar Y_{,r}=\bar V_{,r}\bar X_{,r}=\mathcal{O}(\alpha^2)$. Additionally, we can neglect the terms $\bar X_{,rr}=\bar Y_{,rr}=\bar V_{,rr}=\mathcal{O}(\alpha/r^{k})$, $k\geq 3$, being subdominant with respect to the other terms.}.

Now, let us define $\mathbb{X}\equiv n\mathbb{\mathbb{V}}$, with $n=(l+2)(l-1)/2$. This allows us to recast Eq.~\eqref{eq:03new} as
\begin{equation}\label{eq:sost}
    2\mathbb{T}-l(l+1)\mathbb{V}=-2(\mathbb{L}+\mathbb{X}+\delta\tilde{W})\,,
\end{equation}
which, substituted into Eqs.~\eqref{eq:02new} and \eqref{eq:23new}, gives
\begin{align}
&\left(\mathbb{L}+\mathbb{X}+\frac{\delta\tilde{W}}{2}\right)_{,r}+\left(\frac{1}{r}-\bar\nu_{,r}\right)\left(\mathbb{L}+\mathbb{X}+\frac{\delta\tilde{W}}{2}\right)\nonumber\\
&+\mathbb{L}\left(\frac{1}{r}+\frac{\bar W_{,r}}{2 \bar W}\right)+\frac{1}{2}\left[\delta\tilde W\left(\frac{1}{r}-\frac{ \bar W_{,r}}{ \bar W}\right)+\frac{\tilde{\mathbb{U}}}{r}\right]=0\,, \label{eq:02newnew}
\end{align}
and
\begin{align}
(\mathbb{N}-\mathbb{L})_{,r}=&\left(\frac{1}{r}-\bar\nu_{,r}\right)\mathbb{N}+\left(\frac{1}{r}+\bar\nu_{,r}+\frac{\bar W_{,r}}{\bar W}\right)\mathbb{L}\nonumber\\
&+\delta\tilde{W}\left(\frac{1}{r}-\frac{\bar W_{,r}}{\bar W}\right)+\frac{\tilde{\mathbb{U}}}{r}\,,\label{eq:23newnew} 
\end{align}
respectively. Moreover, taking the difference between Eq.~\eqref{eq:22} and Eq.~\eqref{eq:44}, and considering the terms proportional to $P_{l,\theta}\cot\theta$ only, we find \cite{Chen:2021pxd}
\begin{align}
    &\mathbb{V}_{,rr}+\left(\frac{2}{r}+\bar\nu_{,r}+\bar\mu_{2,r}+\frac{\bar W_{,r}}{\bar W}\right)\mathbb{V}_{,r}+\omega^2 \mathbb{V} e^{-2(\bar\nu+\bar\mu_2)}\notag \\
    &+\frac{e^{-2\bar\mu_2}}{r^2}\left(\mathbb{N}+\mathbb{L}+\delta\tilde{W}\right)=0\,.
    \label{eq:22-44}
\end{align}

Interestingly, the $\delta\tilde{W}$ term in the above expressions can be cancelled out via the redefinitions
\begin{equation}
\mathbb{\hat{L}}\equiv \mathbb{L}+\frac{\delta\tilde{W}}{2}\,,\quad \mathbb{\hat{N}}\equiv \mathbb{N}+\frac{\delta\tilde{W}}{2}\,.
\end{equation}
Hence, Eqs.~\eqref{eq:02newnew}, \eqref{eq:23newnew}, and \eqref{eq:22new} become, respectively,
\begin{subequations}
\begin{align}
    &\left(\hat{\mathbb{L}}+\mathbb{X}\right)_{,r}+\left(\frac{1}{r}-\bar{\nu}_{,r}\right)(\hat{\mathbb{L}}+\mathbb{X})+\hat{\mathbb{L}}\left(\frac{1}{r} +\frac{\bar W_{,r}}{2\bar W}\right)\notag\\
    &=-\frac{\tilde{\mathbb{U}}}{2r}\,, \label{eq:POL1}\\
    &\left(\hat{\mathbb{N}}-\hat{\mathbb{L}}\right)_{,r}-\left(\frac{1}{r}-\bar\nu_{,r}\right)\hat{\mathbb{N}}-\left(\frac{1}{r}+\bar\nu_{,r}+\frac{\bar W_{,r}}{W}\right)\hat{\mathbb{L}}=\frac{\tilde{\mathbb{U}}}{r}\,, \label{eq:POL2}\\3
    & \frac{1}{r}\left(2+r\frac{\bar W_{,r}}{\bar W}\right)\hat{\mathbb{N}}_{,r}-2\left(\frac{1}{r}+\bar\nu_{,r}+\frac{\bar W_{,r}}{\bar W}\right)\left(\hat{\mathbb{L}}+\mathbb{X}\right)_{,r} \notag \\
    &-\frac{2}{r}\left[\frac{1}{r}+2\bar\nu_{,r}+\left(2+r\bar\nu_{,r}\right)\frac{\bar W_{,r}}{\bar W}\right]\hat{\mathbb{L}}-\frac{l(l+1)}{r^2}\hat{\mathbb{N}}\, e^{-2\bar\mu_2}\notag \\
    &+\frac{2n}{r^2}\left(\hat{\mathbb{L}}-\mathbb{V}\right)e^{-2\bar\mu_2}-2\omega^2\left(\hat{\mathbb{L}}+\mathbb{X}\right)e^{-2(\bar\nu+\bar\mu_2)}\notag \\
    &=-\left(2-e^{-2\bar\mu_2}\right)\left(\frac{\tilde{\mathbb{U}}}{r}\right)_{,r}\,, \label{eq:POL3}
\end{align}
\end{subequations}
where we have neglected the terms $\delta\tilde W\frac{\bar W_{,r}}{\bar W}$ and $\delta\tilde{W} _{,r}\frac{\bar W_{,r}}{\bar W}$, as they are second-order perturbations.

Substituting Eq.~\eqref{eq:POL1} into Eq.~\eqref{eq:POL3}, one has
\begin{align}
&\left(1+\frac{r \bar W_{,r}}{2\bar W}\right)\hat{\mathbb{N}}_{,r}=a_1 \hat{\mathbb{N}}+a_2 \hat{\mathbb{L}}+ a_3 \mathbb{X}-a_4\,,
\label{eq:POL1new}
\end{align}
while, substituting Eq.~\eqref{eq:POL1new} into Eq.~\eqref{eq:POL2}, we obtain
\begin{align}
\tilde{\mathbb{L}}_{,r}=&\left(\bar\nu_{,r}-\frac{1}{r}+\frac{2 a_1 \bar W}{2 \bar W+r \bar W_{,r}}\right)\hat{\mathbb{N}}\notag\\
&+\left(\frac{2 a_2  \bar W}{2\bar W+r \bar W_{,r}}-\frac{\bar W_{,r}}{\bar W}-\frac{1}{r}-\bar\nu_{,r}\right)\hat{\mathbb{L}}\notag\\
&+\left(\frac{2 a_3  \bar W}{2 \bar W+r \bar W_{,r}}\right) \mathbb{X}-\frac{2a_4 \bar W}{2\bar W+r\bar W_{,r}}-\frac{\tilde{\mathbb{U}}}{r}\,.\label{eq:POL2new}
\end{align}    
Finally, plugging Eq.~\eqref{eq:POL2new} into Eq.~\eqref{eq:POL1} yields
\begin{align}
\mathbb{X}_{,r}=&\left(\frac{1}{r}-\bar\nu_{,r}-\frac{2 a_1 \bar W}{2 \bar W+r \bar W_{,r}}\right)\hat{\mathbb{N}}\notag\\
&+\left(2\bar\nu_{,r}-\frac{1}{r}+\frac{\bar W_{,r}}{2\bar W}-\frac{2 a_2  \bar W}{2\bar W+r \bar W_{,r}}\right)\hat{\mathbb{L}}\nonumber\\
&+\left(\bar\nu_{,r}-\frac{1}{r}-\frac{2 a_3  \bar W}{2 \bar W+r \bar W_{,r}}\right)\mathbb{X}+\frac{2a_4\bar W}{2\bar W+r\bar W_{,r}}+\frac{\tilde{\mathbb{U}}}{2r}\,,
\label{eq:POL3new}
\end{align}    
where we have defined
\begin{subequations}
\begin{align}
a_1\equiv &\ \left(\frac{n+1}{r}\right) e^{-2 \bar\mu_2}\,,\\
a_2\equiv &\ e^{-2 \bar \mu_2} \left(e^{-2 \bar \nu} r \omega ^2-\frac{n}{r}\right)-\frac{\bar W_{,r} }{2 \bar W}\left(1-3 r \bar \nu_{,r}\right)\notag\\
&+\bar \nu_{,r} \left(1+r \bar \nu_{,r}\right)-\frac{r}{2}{\left(\frac{ \bar W_{,r}}{\bar W}\right)}^2-\frac{1}{r}\,,\\
a_3\equiv &\ e^{-2 \bar\mu_2} \left(\frac{1}{r}+r \omega ^2e^{-2 \bar \nu} \right)-\frac{\bar W_{,r}}{\bar W}\left(1-r \bar \nu_{,r}\right)+r \bar \nu_{,r}^2-\frac{1}{r}\,, \\
a_4 \equiv &\ \frac{r}{2} \left[\left(\frac{1}{r}+\bar\nu_{,r}+\frac{ \bar W_{,r}}{\bar W}\right)\frac{\tilde{\mathbb{U}}}{r} + \left(2-e^{-2 \bar\mu_2}\right) \left(\frac{\tilde{\mathbb{U}}}{r}\right)_{,r}\right].
\end{align}    
\end{subequations}

Using the tortoise coordinate~\eqref{eq:tortoise}, it is possible to recast the differential problem into the wave-like form
\begin{equation}
\Psi_{,r_*r_*}^{(\text{pol})}+\left[\omega^2+V^{(\text{pol})}\right]\Psi^{(\text{pol})}= \lambda_0 \tilde{\mathbb{U}}+\lambda_1 \tilde{\mathbb{U}}_{,r}+\lambda_2 \tilde{\mathbb{U}}_{,rr}\,,
    \label{eq:Schrodinger}
\end{equation}
where 
\begin{subequations}
\begin{align}
\lambda_0&\equiv\frac{1}{2 n r^3 (n r+3)^2}\Big\{n^2 r^6 \omega ^2+4 (n-2) n^2 r^3+6 n r^5 \omega ^2\notag\\
&+r^4 \left(-2 n^3-2 n^2+9 \omega ^2\right)+14 n (2 n-3) r^2+108\notag\\
&+36 (3 n-1) r\Big\}\,,\\
\lambda_1&\equiv\frac{n \left(r^2-5\right) r+3 (r-1) r-9}{n r^2 (n r+3)}\,,\\
\lambda_2&\equiv\frac{(r-4) (r-2)}{2 n r}\,.
\end{align}
\end{subequations}
Moreover, we have $\Psi^{(\text{pol})}=\Psi_0^{(\text{pol})}+\alpha\Psi_1^{(\text{pol})}$ with
\begin{subequations}
    \begin{align}
    &\Psi_0^{(\text{pol})}= \frac{r^2}{nr+3}\left(\frac{3\mathbb{X}}{nr}-\hat{\mathbb{L}}\right), \\
    &\Psi_1^{(\text{pol})}= \frac{(r-3)^{-1}}{2(nr+3)^2}\!\Bigg\{\!\frac{9r\big[r^{1 - k} + 3^{-k}\big(n(r-3)r-3\big)\big]}{n}\mathbb{X}\notag \\
    &\hspace{0.4cm} +r^2 \Big[3^{1-k} \big((n+3) r-3\big)-r^{1-k} (n r+6)\Big]\hat{\mathbb{L}}\Bigg\}.
    \end{align}
\end{subequations}
Likewise, the potential $V^{(\text{pol})}$ takes the form 
\begin{equation}
    V^{(\text{pol})}=V_0^{(\text{pol})}+\alpha V_1^{(\text{pol})}\,,
\end{equation}
where 
\begin{subequations}
\begin{align}
    &V_0^{(\text{pol})}= \frac{2 (r-2)\left[n^2 r^2 (n r + r+3) + 9 (n r+1)\right]}{r^4 (n r+3)^2}\,, \\
    &V_1^{(\text{pol})}= -\frac{2\left[n^2 r^2 (n r + r+3) + 9 (n r+1)\right]}{r^{3+k}(n r+3)^2}\notag\\
    &+\frac{3(r-2)(r-3^{1-k}r^k)\left[n^2 r^2 (n r + 2 r - 3) - 9 ( n r + 1)\right]}{r^{4+k}(r-3) (n r+3)^3}.
\end{align}
\end{subequations}
It is worth noting that $V_0^{(\text{pol})}$ coincides with the Zerilli potential \cite{Zerilli:1970wzz}.

In order to solve Eq.~\eqref{eq:Schrodinger}, we need an independent equation for $\mathbb{U}$. Specifically, in our case, we may use $\Box\mathbb{U}=0$  (see Appendix~\ref{app:4.41_CP} for the derivation details).
From the first line of Eq.~\eqref{U}, one then finds \cite{Chen:2021pxd}
\begin{equation}\label{eq:Schroedinger-U}
\tilde{\mathbb{U}}_{,r_*r_*}+\left(\omega^2-V_{\mathbb{U}}\right)\tilde{\mathbb{U}}=0\,,
\end{equation}
where 
\begin{equation}
V_{\mathbb{U}}=\frac{2 (r-2)}{r^4}\big[(n+1)r+1)\big]\,.
\end{equation}
Expressing $\tilde{\mathbb{U}}_{,rr}$ from Eq.~\eqref{eq:Schroedinger-U}, Eq.~\eqref{eq:Schrodinger} simplifies to
\begin{equation}
        \Psi_{,r_*r_*}+\left[\omega^2+V^{\text{(pol)}}\right]\Psi= \tilde{\lambda}_0 \tilde{\mathbb{U}}+\tilde{\lambda}_1 \tilde{\mathbb{U}}_{,r}\,,
    \label{eq:Schrodinger2}
\end{equation}
where
\begin{subequations}
\begin{align}
\tilde{\lambda}_0&\equiv\frac{1}{n r^3}\bigg\{\frac{n r [6 (r-5) r-6-n r (2 n r+r+14)]}{(n r+3)^2}\notag \\
&+\frac{r^4 \omega ^2}{r-2}+\frac{9 (r-5) r+18}{(n r+3)^2}\bigg\}\,,\\
\tilde{\lambda}_1&\equiv\frac{n [(r-1) r-1] r+3 (r-2) r+3}{n r^2 (n r+3)}\,.
\end{align}    
\end{subequations}

These results show that, in contrast with GR, the evolution of polar perturbations is also influenced by the dynamics of the auxiliary field $\mathbb{U}$. The nonlocal contributions enter both as a correction to the Zerilli potential and through the additional source terms in Eq.~\eqref{eq:Schrodinger2}.

\section{Conclusions}
\label{sec:conclusions}

In this work, we have developed a theoretical framework to describe gravitational perturbations of a static and spherically symmetric spacetime within the revised DW theory of gravity. In particular, we have focused on the DD BH metric, which constitutes a family of first-order corrections to the Schwarzschild solution.

We started by considering axial perturbations. In doing so, we analytically demonstrated that the propagation of axial modes of the DD BH, at zeroth-order, reduces to the standard Regge-Wheeler equation. At first order, however, the corrections arising from the nonlocal gravitational terms preclude the resulting differential problem from being expressed in a Schr\"odinger-like form. 

We then extended the analysis to polar perturbations. Through suitable mathematical manipulations, we were able to recast the differential problem into a wave equation that, at zeroth order, recovers the Zerilli equation. Furthermore, we showed that the presence of nonlocal-sourced terms gives rise to a modified Schr\"odinger-like equation, governed by a characteristic nonlocal function that, in turn, satisfies a wave-type equation.
It should be emphasized that, although the present study is entirely theoretical and involves technically demanding calculations, its generality ensures that our method can be directly applied or extended to any static and spherically symmetric BH background within the considered class of nonlocal gravity theories. 

A natural and physically relevant extension of this work would consist of computing the associated QNM spectrum, following the approach adopted in our previous study \cite{DAgostino:2025yej}, but now incorporating the full set of axial and polar perturbations derived here. From a technical perspective, this represents a significant challenge, as the perturbation equations do not generally reduce to a Schrödinger-like form. As a consequence, many of the standard techniques available in the literature cannot be directly employed, thus requiring the development and implementation of new analytical or numerical methods. 

Besides exploiting GW signals from the ringdown phase of binary BH mergers to test nonlocal effects, complementary studies can be performed in the strong-field regime through electromagnetic observations of astrophysical phenomena. In this framework, the Event Horizon Telescope is particularly relevant, as it enables direct probes of near-horizon physics~\cite{Ayzenberg:2023hfw}. One should focus on the region extending from the photosphere down to the event horizon, where gravitational fields are strongest. Valuable information can be extracted from high-resolution imaging and BH shadow measurements, as well as from analyses of accretion dynamics and infalling matter. A comprehensive set of electromagnetic-based strategies is presented in Ref.~\cite{DeFalco:2023rxm}, where observables derived from spacetimes slightly departing from the Schwarzschild metric are discussed.

Nevertheless, it is not possible to identify \emph{a priori} which electromagnetic observables are most promising for current or near-future experiments, since this cannot be inferred directly from the metric itself. As noted for the GW sector, it is essential to perform explicit calculations to determine how nonlocal effects manifest observationally and, subsequently, to develop strategies that may amplify these signatures and render them detectable. We leave this task to future investigations.

\acknowledgements
The authors would like to thank the anonymous referee for interesting comments and suggestions.
RD thanks the Istituto Nazionale di Fisica Nucleare (INFN), Sezione di Roma 1, \textit{esperimento} Euclid, for financial support. RD acknowledges work from COST Action CA21136 -- Addressing observational tensions in cosmology with systematics and fundamental physics (CosmoVerse).
VDF is grateful to Gruppo Nazionale di Fisica Matematica of Istituto Nazionale di Alta Matematica (INDAM) for support. VDF acknowledges the support of INFN, Sezione di Napoli, {\it iniziativa specifica} TEONGRAV.

\appendix
\section{Linearized field equations for the Schwarzschild background}
\label{sec:Sch-FE}
Here, we compute the linear perturbations of the field equations~\eqref{eq:linearized} assuming the Schwarzschild spacetime as the background metric. In this case, we have $-g_{tt}=1/g_{rr}=e^{2\bar \nu}=1-2/r$ and $\bar W$ is a constant.
Our results may be directly compared with those obtained in Ref.~\cite{Chen:2021pxd} by applying the following substitutions for the auxiliary scalar fields: $\left\{X\mapsto\phi,\ U\mapsto-\Xi, \ V\mapsto-\psi,\ Y\mapsto \zeta,\ W\mapsto 1+K\right\}$.

Specifically, the $(t,r), (t,\theta)$, and $(r,\theta)$ components are, respectively:
\begin{subequations}
\begin{align}
&\bar W\left[\left(\delta\psi+\delta\mu_3\right)_{,r}+\left(\frac{1}{r}-\bar\nu_{,r}\right)\left(\delta\psi+\delta\mu_3\right)-\frac{2}{r}\delta\mu_2\right]\nonumber\\
&+e^{\bar\nu}\left(e^{-\bar\nu}\delta W\right)_{,r}-\frac{1}{2}\big(\bar X_{,r}\delta U+\bar U_{,r}\delta X+\bar V_{,r}\delta Y \nonumber\\
&+\bar Y_{,r}\delta V+2\bar V\bar X_{,r}\delta X\big)=0\,,
\label{02sc}
\\
 &\bar W\Big[(\delta\psi+\delta\mu_2)_{,\theta}+(\delta\psi-\delta\mu_3)\cot\theta\Big]\!+\delta W_{,\theta}=0\,,
 \label{03sc}
\\
&\bar W\bigg[\left(\delta\psi+\delta\nu\right)_{,r\theta}+\left(\delta\psi-\delta\mu_3\right)_{,r}\cot\theta +\left(\bar\nu_{,r}-\frac{1}{r}\right)\delta\nu_{,\theta}  \nonumber\\
&-\left(\bar\nu_{,r}+\frac{1}{r}\right)\delta\mu_{2,\theta}\bigg]= \frac{1}{r}\delta W_{,\theta}-\delta W_{,r\theta} +\frac{1}{2}\big(\bar X_{,r}\delta U_{,\theta} \notag \\
&+\bar U_{,r}\delta X_{,\theta}+\bar V_{,r}\delta Y_{,\theta}+\bar Y_{,r}\delta V_{,\theta}+2\bar V\bar X_{,r}\delta X_{,\theta}\big)\,.
\label{23sc}
\end{align}
\end{subequations}
Moreover, for the $(r,r)$ and $(\varphi,\varphi)$ components we obtain, respectively:
\begin{subequations}
\begin{align}
    &e^{2\bar\nu}\left\{\frac{2}{r}\left[\delta\nu_{,r}-\!\left(\!\frac{1}{r}+2\bar\nu_{,r}\!\right)\!\delta\mu_2\right]+\left(\!\frac{1}{r}+\bar\nu_{,r}\!\right)\!\left(\delta\psi+\delta\mu_3\right)_{,r}\!\right\}\notag \\
    &+\frac{1}{r^2}\left[2\delta\mu_3+\left(\delta\nu+\delta\psi\right)_{,\theta\theta}+\left(2\delta\psi+\delta\nu-\delta\mu_3\right)_{,\theta}\cot\theta\right]\notag \\
    &+e^{-2\bar\nu}\omega^2(\delta\psi+\delta\mu_3)=\frac{-1}{\bar W}\bigg\{
    \frac{1}{r^2}\left(\delta W_{\theta\theta}+\delta W_{,\theta}\cot\theta-\delta W\right)\notag \\
    &+e^{2\bar\nu}\left[\left(\frac{1}{r}+2\bar\nu_{,r}\right)\frac{\delta W}{r}+\left(\frac{2}{r}+\bar\nu_{,r}\right)\delta W_{,r}\right]+e^{-2\bar\nu}\omega^2\delta W \notag\\
    &+\!\left(\frac{1}{2}-e^{2\bar\nu}\!\right)\!\left(\bar X_{,r}\delta U_{,r}+U_{,r}\delta X_{,r}+\bar V_{,r}\delta Y_{,r}+\bar Y_{,r}\delta V_{,r}\right.\nonumber\\
    &\left.+\bar X_{,r}^2\delta V+2\bar V\bar X_{,r}\delta X_{,r}\right)\!\bigg\}\,,
    \label{22sc}
    \\
    &e^{2\bar\nu}\bigg[\left(\delta\nu+\delta\mu_3\right)_{,rr}+\left(\frac{1}{r}+\bar\nu_{,r}\right)\left(2\delta\mu_3-\delta\mu_2\right)_{,r}\notag\\
    &+\!\left(\frac{1}{r}+3\bar\nu_{,r}\!\right)\!\delta\nu_{,r}\bigg]\!+\frac{1}{r^2}(\delta\nu+\delta\mu_2)_{,\theta\theta}+e^{-2\nu}\omega^2(\delta\mu_2+\delta\mu_3)\nonumber\\
    &=\frac{-1}{\bar W}\bigg\{e^{-2\bar\nu}\omega^2\delta W+e^{2\bar\nu}\left[\delta W_{,r}\left(\frac{1}{r}+2\bar\nu_{,r}\right)+\delta W_{,rr}\right]\notag \\
    &+\frac{\delta W_{,\theta\theta}}{r^2}-\frac{1}{2}(\bar X_{,r}\delta U_{,r}+\bar U_{,r}\delta X_{,r}+\bar V_{,r}\delta Y_{,r}+\bar Y_{,r}\delta V_{,r}\nonumber\\
    &+\bar X_{,r}^2\delta V+2\bar V\bar X_{,r}\delta X_{,r})\bigg\}\,.
    \label{44sc}
\end{align}
\end{subequations}

A close comparison of Eqs.~\eqref{22sc} and \eqref{44sc} with Eqs.~(4.18) and (4.17) of Ref.~\cite{Chen:2021pxd}, respectively, reveals that the latter contain some inaccuracies, mainly in the terms involving the auxiliary scalar fields.

\section{Nonlocal-sourced terms of the axial wave equation}
\label{app:axial}

The first-order correction to the effective potential of the axial perturbations reads

\begin{align}
    &V_1^{(\text{ax})}= \frac{3^{-k} r^{-k-4}}{2 (r-3)^4 (r-2)^2}\bigg\{3^k r \Big[2 k^3 \left(r^2-5 r+6\right)^3 r^4\notag \\
    &-k^2 \left(r^2-5 r+6\right)^2 \big(10 r^6-74 r^5+114 r^4-r^3+7 r^2\notag \\
    &-16 r+12\big)+k \big(24 r^{10}-408 r^9+2978 r^8-11665 r^7\notag \\
    &+25471 r^6-29025 r^5+12651 r^4+2626 r^3-3540 r^2\notag \\
    &+2664 r-864\big)+2 \Big((18-28 n) r^9+2 (78 n-125) r^8 \notag \\
    &+(1594-432 n) r^7+11 (54 n-467) r^6+(7795-324 n) r^5\notag\\
    &-2994 r^4-5832 r^3+9570 r^2-8568 r+3240+2 n r^{10}\Big)\Big]\notag \\
    &-6 r^k\big(6 r^{10}-78 r^9+468 r^8-1435 r^7+2089 r^6-750 r^5 \notag \\
    &-1882 r^4+4174 r^3-5556 r^2+4104 r-1296\big) \bigg\}\,.
     \label{eq:V1_axial}
\end{align}

Moreover, we report in the following the full expressions of the coefficients $\beta_1$ and $\beta_2$ in the right-hand side of Eq.~\eqref{eq:AXIAL}:
\begin{subequations}
\begin{align}
    \beta_1=&\ \frac{3^{-k} r^{-k-4}}{2 (r-3)^3 (r-2)} 
    \bigg\{3^k r^3 \Big[5r^6(k^2-3 k+4)  \notag \\
    &-10r^5(5 k^2-19 k+24) +r^4(185 k^2-845 k+1162)  \notag \\
    &-r^3(299 k^2-1593 k+2476) +r(57 k^2+137 k-468)  \notag \\
    &+2r^2(84 k^2-556 k+991) -2 (67 k^2+148 k-690)\Big]\notag\\
    &+4r\,3^{1 + k} \Big[90 - 163 r + 3 k\left(9r-4\right) + k^2 \left(13r-6\right)\Big]  \notag\\
    &-6r^k \big(6 r^8-55 r^7+242 r^6-505 r^5+344 r^4+220 r^3\notag \\
    &-470 r^2+504 r-216\big)\bigg\}\,, \label{eq:beta1}
    \\
    \beta_2=&\ \frac{r^4+r-2}{r^{3+k}(r-3)^2}\Big\{3^{1-k}r^k  (r-2) (2 r-3) \notag \\
    &-r \big[k (r-3) (r-2)-(r-10) r-18\big]\Big\}\,. \label{eq:beta2}
\end{align}
\end{subequations}

\section{Proof of $\Box\mathbb{U}=0$}
\label{app:4.41_CP}
Taking into account Eqs.~\eqref{eq:X_sol}--\eqref{eq:W_sol}, Eq.~\eqref{eq:fY}, and \eqref{eq:Ricci}, we can conveniently recast the scalar fields as
\begin{equation}
    \bar{X}=\alpha X_1\,,\  \bar{Y}=\alpha Y_1\,, \ \bar W=1+\alpha W_1\,, \ f=\alpha f_1\,, \ \bar R=\alpha R_1\,,
    \label{eq:scalar_dec}
\end{equation}
so that, Eq.~\eqref{U} becomes
\begin{align}
    \mathbb{U}&=\frac{\alpha\left(X_1'\delta U+V_1'\delta Y+Y_1'\delta V\right)+2\alpha^2V_1X_1'\delta X}{2(1+\alpha W_1)}\,.
    \label{eq:U_series}
\end{align}
Moreover, using the geometric series expansion
\begin{equation}
    \frac{1}{2(1+\alpha W_1)}=\frac{1}{2}\left[1-\alpha W_1+\mathcal{O}(\alpha^2)\right]\,,
\end{equation}
we can rewrite Eq.~\eqref{eq:U_series} as
\begin{equation}
    \mathbb{U}=\frac{\alpha}{2}\left(X_{1,r}\delta U+V_{1,r}\delta Y+Y_{1,r}\delta V\right)+\mathcal{O}(\alpha^2)\,.
    \label{eq:U_series_2}
\end{equation}
From Eq.~\eqref{eq:X}, we have
\begin{equation}
    \Box \bar X+\Box\delta X=\bar R+\delta R\,.
\end{equation}
Thus, using $\Box\bar X=\delta R$, we find
\begin{equation}
    \Box \delta X=\delta R\,.
     \label{eq:box_deltaX}
\end{equation}
Additionally, from Eq.~\eqref{eq:Y}, we have
\begin{align}
    \Box \bar Y+\Box \delta Y& =g^{\mu\nu}\nabla_\mu(\bar X+\delta X)\nabla_\nu(\bar X+\delta X)\notag \\
    &=(\partial \bar X)^2+2(\nabla^\mu \bar X)(\nabla_\nu \delta X)+\mathcal{O}(\delta^2)\,.
\end{align}
Since $\Box \bar{Y}=(\partial \bar X)^2$, we get
\begin{equation}
    \Box\delta Y=2\alpha(\nabla^\mu X_1)(\nabla_\mu\delta X)\,. 
    \label{eq:box_deltaY}
\end{equation}
Then, Eq.~\eqref{eq:U} reads
\begin{align}
    &\Box U+\Box \delta U=-2\nabla_\mu\left[(\bar V+\delta V)\nabla^\mu(\bar X+\delta X)\right]\notag \\
&=-2\nabla_\mu(\bar V\nabla^\mu\bar X)-2\nabla_\mu(\bar V\nabla^\mu\delta X)-2\nabla_\mu(\delta V\nabla^\mu \bar X)\notag \\
&+\mathcal{O}(\delta^2)\,.
\end{align}
Using $\Box\bar U=-2\nabla_\mu(\bar V\nabla^\mu\bar V)$, we get
\begin{align}
    \Box\delta U=-2\alpha\nabla_\mu(V_1\nabla^\mu\delta X+\delta V\nabla^\mu X_1)\,. 
    \label{eq:box_deltaU}
\end{align}
Finally, from Eq.~\eqref{eq:V}, we have
\begin{equation}
    \Box\bar V+\Box\delta V=-(\bar R+\delta R)\left[f_{,Y}(\bar Y)+f_{,YY}(\bar Y)\delta Y+\mathcal{O}(\delta^2)\right],
\end{equation}
Since $\Box\bar\psi=-\bar Rf'(\bar\zeta)$ and neglecting the terms of orders $\mathcal{O}(\alpha^2,\delta^2)$, eventually we find 
\begin{equation}
    \Box\delta V=\alpha f_{1,Y}\delta R\,.
    \label{eq:box_deltaV}
\end{equation}
Therefore, using Eqs.~\eqref{eq:box_deltaX}, \eqref{eq:box_deltaY}, \eqref{eq:box_deltaU}, and \eqref{eq:box_deltaV} to compute Eq.~\eqref{eq:U_series_2} at first-order in $\alpha$, we conclude that 
\begin{align} \label{eq:boxU}
    \Box\mathbb{U}&=\frac{1}{2}\left(\delta U\,\Box \bar X_{,r}+\delta Y\,\Box \bar V_{,r}+\delta V\, \Box \bar Y_{,r}\right)\simeq 0\,,
\end{align}
where in the last step we have employed the assumptions already considered in Eq.~\eqref{eq:rel_scalar_2}.

%


\end{document}